\documentclass[prl,reprint,superscriptaddress]{revtex4-1}
\usepackage{mathtools}
\usepackage{amsmath,amssymb}
\usepackage{color}
\usepackage{braket}
\usepackage{graphicx}
\usepackage{epstopdf}
\usepackage{float}
\usepackage{nccmath}
\usepackage{times}
\usepackage[normalem]{ulem} 
\usepackage{textgreek}
\usepackage{colortbl}
\usepackage{multirow}
\usepackage{hhline}

\usepackage{soul}
\usepackage{hyperref}
\usepackage[table]{xcolor}

\begin{document}


\title{Stretched-Exponential Melting of a Dynamically Frozen State Under Imprinted Phase Noise in the Ising Chain in a Transverse Field}

\author{Krishanu Roychowdhury}
\email[]{krishanu.1987@gmail.com}
\affiliation{Theory Division, Saha Institute of Nuclear Physics, HBNI, 1/AF Bidhannagar, Kolkata 700064, India.}
\affiliation{Max-Planck-Institut f\"{u}r Physik komplexer Systeme, N\"{o}thnitzer Strasse 38, Dresden 01187, Germany.}

\author{Arnab Das}
\email[]{arnab.das.physics@gmail.com}
\affiliation{Indian Association for the Cultivation of Science, 2A \& 2B Raja S. C. Mullick Road, Kolkata 700032, India.}

\begin{abstract}
Dynamical freezing is a phenomenon where a set of local observables emerges as approximate but stable conserved quantities (freezes) under a strong periodic drive in a closed quantum system. The expectation values of these emergent conserved quantities exhibit small fluctuations around their respective initial values. These fluctuations do not grow with time, and their magnitude can be tuned down sharply by tuning the drive parameters. In this work, we probe the resilience of dynamical freezing to random perturbations added to the relative phases between the interfering states (elements of a natural basis) in the time-evolving wave function after each drive cycle. We study this in an integrable Ising chain in a time-periodic transverse field. Our key finding is, that the imprinted phase noise melts the dynamically frozen state, but the decay is ``slow": a stretched-exponential decay rather than an exponential one. Stretched-exponential decays (also known as Kohlrausch relaxation) are usually expected in complex systems with time-scale hierarchies due to strong disorders or other inhomogeneities resulting in jamming, glassiness, or localization.
\end{abstract}

\maketitle


\noindent
{\bf Introduction:} The study of non-equilibrium phenomena in driven quantum systems has sparked significant interest \cite{Anatoli_Krish_RMP, bukov2015universal, Bloch_Gross_Rev, Blatt_Monroe_Wineland_RMP, Eisert_Gogolin_Rev, Haldar_Rev, Ho_Mori_Abanin}, uncovering phenomena far from the realm of the notions and concepts of equilibrium physics or its intuitive extensions even in ostensibly simple tractable systems. In this work we focus on one such example - the phenomenon of {\it dynamical freezing} (DF) \cite{das2010exotic, bhattacharyya2012transverse, hegde2014freezing, Onset, haldar2021dynamical, KSG_Freezing, russomanno2012periodic, roy2015fate, Naveen_Freezing, bukov2015universal, Haldar_Rev, Eckstein_Rev}, in particular, an incarnation of it in an integrable spin chain. In DF, a strongly driven closed quantum system (including quantum-chaotic interacting many-body systems) can evade the intuitive Floquet thermalization to a locally infinite-temperature-like state~\cite{lazarides2014equilibrium,d2014long}, when the drive amplitude crosses a threshold~\cite{Onset}. This happens due to the appearance of exact (in systems mappable to non-interacting fermions or hard-core bosons) \cite{lazarides2014periodic} and emergent (in generic interacting systems) \cite{das2010exotic, Onset, haldar2021dynamical} stroboscopic conservation laws that are not present in the undriven system. Here we investigate the robustness of the phenomenon in an integrable spin chain, against adding after each drive cycle a random component to the relative phases between the amplitudes of the members of a particular (natural) basis in the wave function. Our study reveals a rich pallet of interesting regimes of stability and decay. Previously, DF was argued to result from repeated quantum interference of the relative phases mentioned above \cite{hegde2014freezing, Haldar_Rev}. In particular, it was shown earlier by Amit Dutta and collaborators that if one considers the dynamics discounting the phases altogether, i.e., consider only the {\it transition probabilities} for each cycle ignoring the interference between the {\it transition amplitudes}, then an infinite-temperature-like scenario (within the integrable structure) is approached exponentially rapidly with the number of drive cycles \cite{Amit_Da_Probability}. 

Here we examine the robustness of DF by targeting those phases and corrupting them randomly in a controlled manner. The phenomenology we uncover here is expected to be realizable in experimental setups similar to that described in Ref.~\onlinecite{hegde2014freezing}.


\noindent
{\bf The Model and The Drive Protocol:} The model under investigation is specified by the Hamiltonian 
\begin{align}\label{ham1}
 {\cal H}(h) = -J \sum_{j=0}^{L-1} \big[ \sigma^x_j \sigma^x_{j+1} + h(t)\sigma^z_j \big],
\end{align}
where the Pauli matrices satisfy periodic boundary condition $\sigma^\alpha_L=\sigma^\alpha_0$ for $\alpha \in \{x,z\}$, and admits a ferromagnetic (paramagnetic) phase for $|h|<1$ ($|h|>1$) and critical points at $|h|=1,$ and we set $J=1$.
In our periodic driving protocol, the transverse field is time-dependent, quenched from $+h_0$ to $-h_0$ ($h_0>0$), and back periodically over a period of $T$ such that 
\begin{align}
h(t)= 
\begin{cases}
    +h_0, & \text{for~~} nT < t < (n+1/2)T \\
    -h_0, & \text{for~~} (n+1/2)T < t < (n+1)T\,.
\end{cases}
\end{align}
We follow the system's dynamics stroboscopically at instants $t=nT,$ where $T$ is the period of one drive cycle, and $n$ is a non-negative integer. This sets the final time $t_f=N_c T$ ($n=0,1,\dots, N_c$). The system is prepared to be at the ground state of the Hamiltonian at $t=n=0$, then allowed to evolve under the protocol mentioned above. Measurements of the relevant observables are recorded after $N_c$ number of full cycles.

The standard steps to diagonalize the Hamiltonian at any instant are as follows \cite{suzuki2012quantum}. First, the Pauli spin operators are expressed in terms of (spinless) Jordan-Wigner fermions; the resultant fermionic Hamiltonian reads
\begin{align}\label{ham2}
 {\cal H}(h) = - \sum_{j=0}^{L-1} \big[ (c^\dagger_j-c_j)(c^\dagger_{j+1}+c_{j+1}) + h\sum_{j=0}^{L-1} (1-2c^\dagger_j c_j) \big].
\end{align}
This is a quadratic problem in terms of the fermions, and invoking translation invariance (under periodic boundary condition) can easily be solved in the Fourier space following the Fourier transform $c_k = 1/\sqrt{L} \sum_j c_j e^{-ikj},$ where $k$ denotes the Fourier space quantum numbers with $k\in(-\pi,\pi)$. A particular advantage is that the Hamiltonian in the Fourier space is a non-interacting ensemble of two-level systems ${\cal H}(h) = \bigoplus_k H_k (h) $ with
\begin{align}\label{ham3}
  H_k(h) &= (h-\cos{k}) ~ (c^\dagger_{k}c_{k} -c_{-k}c^\dagger_{-k}) - i\sin{k}~ c^\dagger_{k} c^\dagger_{-k} \nonumber \\
  &~~~~~~~~~~~~~~~~~~~~~~~~~~~~~~~~~~~~~~~~~~~~~~~~~~~~~ + i\sin{k}~ c_{-k} c_{k}.
\end{align}
The manifold in which the ground state and states reachable from it via the Schr\"{o}dinger dynamics under $H_{k}(t)$ lie, is spanned by the set of basis states $|0_k,0_{-k}\rangle$ and $|+k,-k\rangle$ that represent, respectively, the unoccupied and the doubly occupied states of the $\pm k$ fermions. 

Starting from the ground state, the wavefunction of the system at time $t$ is given by $|\psi(t)\rangle = \bigotimes_k |\psi_k(t)\rangle$ where $|\psi_k(t)\rangle = u_k(t)|0_k,0_{-k}\rangle + v_k(t)|+k,{-k}\rangle$ for some complex amplitudes $u_k$ and $v_k$. The phases of these complex numbers are of pivotal importance in the physics we explore through this work. The allowed values of the quantum number $k$ are set by the fermion parity or equivalently the eigenvalue of the operator ${\cal P}=\prod_{j=0}^{L-1} \sigma_j^z$. We place the system in the even-parity sector (where the ground state is found) by considering $L$ to be even which fixes the quantization of the $k$ values as $k = 2\pi(n+1/2)/L,~~ n=-L/2,\dots, L/2-1$. 

Physical observables that are local in $c_{k}$ and $c_{k}^{\dagger}$ are functions of the amplitudes $u_k(t)$ and $v_k(t)$. For instance, time-dependent transverse magnetization is 
\begin{align}\label{magz1}
    m_z(t) = \frac{1}{L} \sum_j \langle \psi(t) |\sigma^z_j|  \psi(t) \rangle = 1-\frac{2}{L} \sum_k |v_k(t)|^2.
\end{align}
 In addition, of particular interest will be the long-time average of $m_z(t)$~\cite{das2010exotic}:
\begin{align}\label{QQ}
  Q = \lim_{t\rightarrow \infty} t^{-1}\int_0^t {\rm d}t'~m_z(t')\,,
\end{align}
\noindent
which is a measure of the degree of symmetry breaking in the system. 
Since each $k$-mode undergoes the time evolution independently, it renders the time-evolution operator accounting for the full dynamics a simple form ${\cal T} = \bigoplus_k T_k$. In other words, with the initial state specified in a given $k$-sector, the time-evolved state is $|\psi_k(t)\rangle=T_k|\psi_k(t=0)\rangle$ where $T_k$ is the relevant part of the full time-evolution operator $\cal T$ when projected onto the subspace spanned by $|0_{+k},0_{-k}\rangle$ and $|+k,-k\rangle$. With this, we can readily compute $u_k(t), v_k(t)$ as $\begin{bmatrix} u_k(t) & v_k(t) \end{bmatrix}^T=T_k\begin{bmatrix} u_k(t=0) & v_k(t=0) \end{bmatrix}^T$, and hence $m_z(t)$.
The repeated interference between the amplitudes $u_k, v_k$ leads to the remarkable phenomenon of DF. In the following section, we provide a brief account of this and highlight some notable features that will constitute the backbone of subsequent investigations. \\

\begin{figure}
 \centering
  \includegraphics[width=\columnwidth]{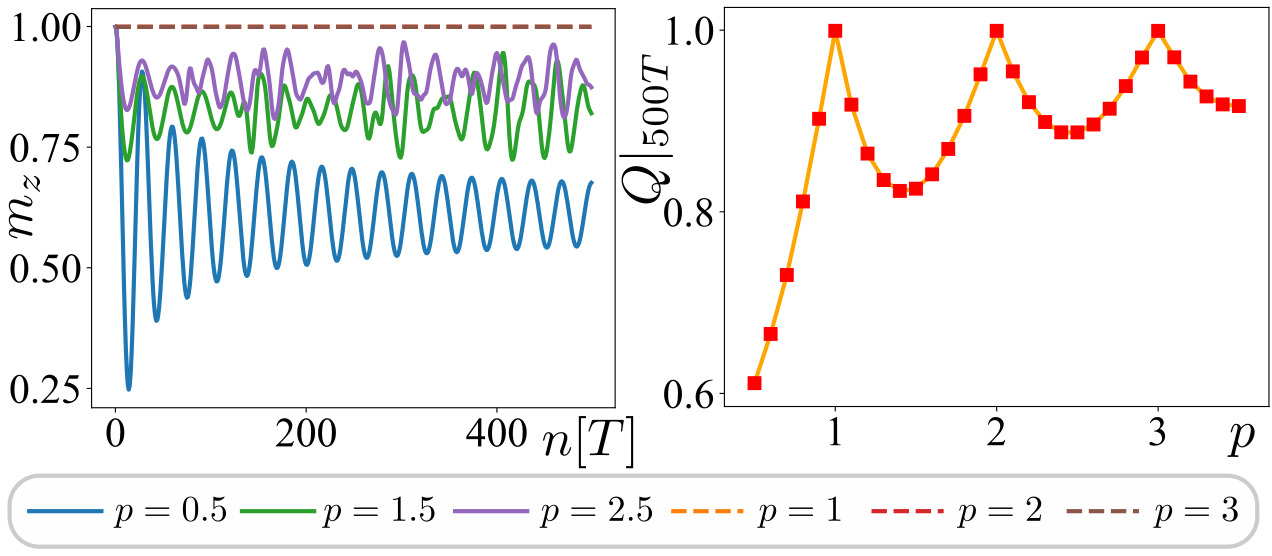}
  \caption{{\bf Left:} Temporal profile of the transverse magnetization $m_z(t)$ at various $p=2h_0/\omega$ with $h_0=20$ plotted over $N_c=500$ full cycles. The time-axis ($t$) is in the units of the full-cycle period $T$ as $t=nT$ where $T=2\pi/\omega$. DF is observed at special frequencies given by the integer values of $p$. {\bf Right:} The long-time averaged magnetization $Q$ after $500$ complete cycles (denoted as $Q|_{500T}$) as a function of $p$ reveals a peak-valley structure where the valleys are located almost halfway between the two consecutive peaks. The results are for the chain with $L=100$.}
 \label{fig1}
\end{figure}


\noindent
{\bf Dynamical Freezing (DF):} For the specific protocol of periodic driving we use, the Hamiltonian remains constant over each half cycle, however, the transverse field flips in the consecutive half cycles. Therefore the time-evolution operator $T_k$ is simply 
\begin{align}
T_k(h)\equiv   
\begin{cases}
    T_k(+h_0) =  e^{-iH_k(h_0) T/2}, & \\~~~~~~~~~~~~~~~~~~~~~~~~~~~~ \text{for~~} nT < t < (n+1/2)T \\
    T_k(-h_0) = e^{-iH_k(-h_0) T/2}, & \\~~~~~~~~~~~~~~~~~~ \text{for~~} (n+1/2)T < t < (n+1)T\,,
\end{cases}
\end{align}
(with $\hbar=1$), and the evolution operator after $N$ full cycles {\it i.e.}, after time $t=NT$ is $U_k(N)=\left[T_k(-h_0)T_k(+h_0)\right]^N$. 

Let us consider the initial state (at $t=0$) to be the ground state of $H_k(h_0)$ at each $k$. It was shown in Ref.~\onlinecite{das2010exotic, bhattacharyya2012transverse} that depending on the driving frequency $\omega =2\pi/T$, the memory of the ${\mathbb Z}_2$ symmetry breaking in the initial state can get imprinted in the response, {\it i.e.}, the transverse magnetization $m_z(t)$ [and so in $Q(t)$] through all later time, though the driving respects this symmetry after each full cycle (over period $T$). This effect is especially prominent when the dimensionless quantity $p=2h_0/\omega$ assumes nonzero integer values -- a condition referred to as the {\it freezing condition}, and the corresponding $\omega$ or the $p$ values, the freezing peaks. 

The profile of $m_z(t=nT)$ as a function of $n$ as well as its long-time average $Q(t=N_cT)$ (denoted as $Q|_{N_cT}$) as a function of $p$ is shown in Fig.~\ref{fig1}. The peaks of $Q$ occur at integer values of $p$, aligning with the initial condition $m_z(t=0)\approx 1$ (set by $h_0=20$), indicating a robust freezing behavior. However, slight deviations from the respective initial values are noticeable at larger integer values of $p$ (increasing with $p$) corresponding to lower-frequency drives which tend towards an approximately adiabatic regime ($\omega\rightarrow 0$ as $p\rightarrow +\infty$). This suggests that the phenomenon of DF is genuinely rooted in nonadiabatic quantum dynamics. 
The plot of $Q|_{500T}$ in Fig.~\ref{fig1} also reveals intriguing valley structures near half-integer values of $p$. These points denote instances where $m_z(t)$ exhibits the largest deviation from its initial values between two consecutive peaks. Remarkably, the magnitude of such deviations diminishes with higher values of the valley locations along the $p$-axis indicating that the valleys at smaller frequencies emulate the freezing behavior of the integer $p$ values at larger frequencies.  

\begin{figure}
 \centering
  \includegraphics[width=\columnwidth]{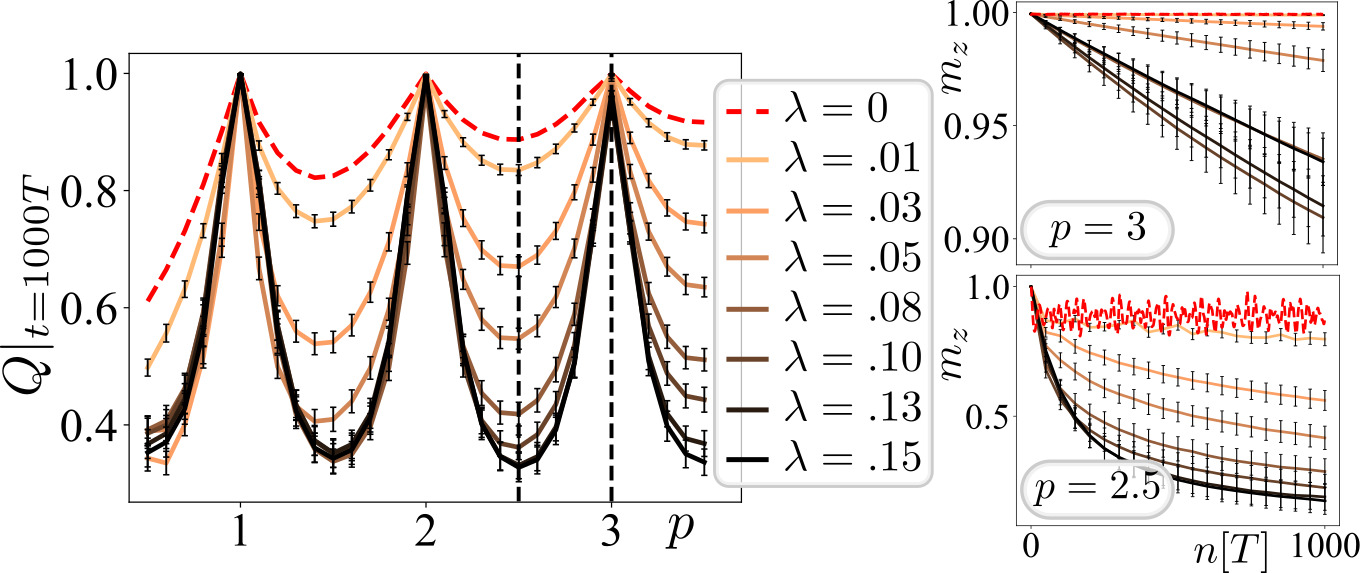}
  \caption{Shown is the peak-valley structure in the behavior of $Q|_{1000T}$ as a function of $p=2h_0/\omega$, in the presence of the random phase noise averaged over $1000$ random realizations (error bars showing the standard deviations) when we vary the strength of the randomness in the added phases $\lambda$. The red dashed line is the plot for $\lambda=0$ (also shown in Fig.~\ref{fig1}). The plots on the {\bf(right)} are for $m_z$ as a function of time along the two vertical dashed lines: the top is at a representative integer value of $p$ (a peak) and the bottom is for a non-integer value (a valley). The red-dashed lines in these again correspond to $\lambda=0$. The results are for the chain with $L=100$.}
 \label{fig2}
\end{figure}

This phenomena of DF cannot be explained in terms of counting the probabilities after each half cycle and the resultant recurrence relation \cite{mukherjee2008defect, hegde2014freezing}. For a specific $k$-mode, the probability of finding it in the excited state in the second half of the $n$-th full cycle reads 
\begin{align}\label{rec1}
   |v_k(n+1/2)|^2 =  |v_k(n)|^2(2P_{\rm ex}(k)-1) + 1-P_{\rm ex}(k),
\end{align}
where we have used the notation $v_k(n+1/2)\equiv v_k(t=(n+1/2)T)$ and $P_{\rm ex}(k)$ is the probability that the state ends up at $|+k,-k\rangle$, following a flip in the transverse field, starting from the same state before the flip. The solution to this relation predicts $|v_k(n+1/2)|^2 \rightarrow 1/2$ independent of $k, T, h_0$ as $n \rightarrow +\infty$ suggesting the transverse magnetization ($m_z$ in Eq.~\ref{magz1}) diminishes over time regardless of $p$. However, the presence of a distinct peak-valley structure in Fig.~\ref{fig1} presents a contrasting scenario; an explanation for DF, therefore, must take into account the amplitudes and their interference rather than solely relying on the probability count outlined above. 

The robustness of the peak-valley structure in this simple integrable model was tested against disorder in the spin-spin interaction as well as in the transverse field \cite{roy2015fate}. Remarkably, it was observed that the decay of the transverse magnetization $m_z(t)$, exponential in time with a characteristic scale $\tau$ (similar to the undriven case), is significantly suppressed at the peak points when plotted against the driving frequency $\omega$. Curiously, this characteristic time scale $\tau(\omega)$ displays an exponential growth with $\omega$ (or inversely with $p$) which warrants that the stability of the peaks intensifies at larger frequencies due to adequate nonadiabatic dynamics induced by the periodic drive even in the presence of disorder. 

Quite intriguingly, the peak-valley structure survives even in scenarios where integrability is compromised. In Ref.~\onlinecite{haldar2021dynamical}, the integrability was broken by introducing higher-neighbor spin-spin interactions, and longitudinal magnetization $m_x(t)$ was measured at late times (in the diagonal ensemble) with the driving field aligning along the ordering direction. When plotted against the driving frequency $\omega$, a peak-valley structure, akin to the integrable counterpart, emerges in the late-time profile of $m_x$ with robust peaks and valleys gaining prominence at larger values of $\omega$.  

These deliberations lay the groundwork for our present inquiry, prompting the question: how does random phase noise in the amplitudes $u_k$ and $v_k$, and their interference, influence the stability of the peak-valley structure within the integrable framework? An extension to the realm of nonintegrable models will be discussed elsewhere. \\


\noindent
{\bf Adding Random Component to The Phases:} To answer this, we set the following protocol of inserting a random phase difference between $u_k$ and $v_k$ after each full cycle and for all $k$ by drawing independent and identically distributed random numbers uniformly over the range $\phi_k(n)\in[0,1]$ (the phenomena we describe here do not alter if the mean of the noise is $0$ {\it i.e.}, $\phi_k(n)\in[-0.5,0.5]$). We then introduce a parameter $\lambda$ as the effective strength of these random phases by modifying $\phi_k(n)\rightarrow \lambda\phi_k(n)$. Such a protocol yields the full time-evolution operator for a specific $k$-mode over a period of $N$ full cycles as $U_k(N)=\prod_{l=1}^N  T_k(-h_0)T_k(+h_0){\cal U}_k(\phi_k(l),\lambda)$, where ${\cal U}_k(\phi_k(l),\lambda)$ is a diagonal matrix of the form ${\cal U}_k(\phi_k(l),\lambda)={\rm diag}\left({\rm Exp}[2i\pi\lambda\phi_k(l)], {\rm Exp}[-2i\pi\lambda\phi_k(l)]\right)$ and $\lambda$ is treated as a control parameter to vary the strength of the random phases as discussed above. In this setting, we embark on analyzing the effect of the interference between such random phases accumulated by the amplitudes after many consecutive cycles on the phenomenon of DF. We aim to elucidate the influence of this interference on both the transverse magnetization $m_z(t)$ and its late-time average $Q$ after a sufficiently large number of complete cycles by scrutinizing the fate of the peak-valley structure.  

\begin{figure}
 \centering
  \includegraphics[width=\columnwidth]{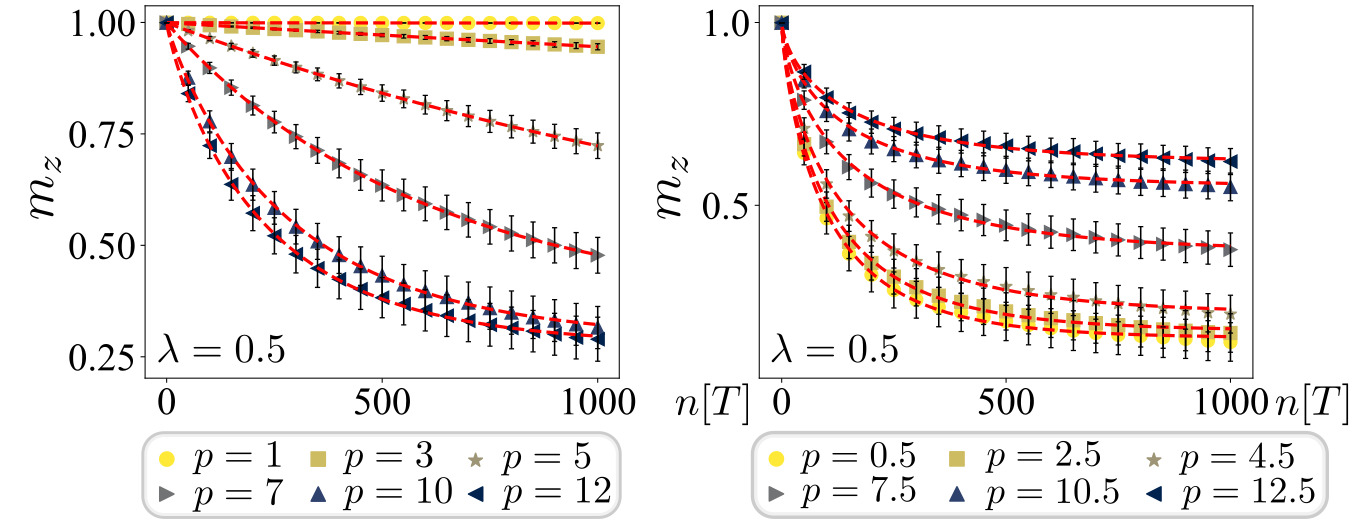}
  \caption{The profile of $m_z(t)$ for various integer values of $p=2h_0/\omega$ {\bf(left)} and non-integer values of $p$ {\bf(right)} at a fixed strength of the phase noise $\lambda=0.5$. The plots fit best to the stretched-exponential of the form $m_z(t) = m_{\rm rm} + (m_0-m_{\rm rm}) {\rm Exp}[-(t/\tau)^\alpha]$ with a $p$-dependent remnant magnetization $m_{\rm rm}$, the steering exponent $\alpha$, and time scale $\tau$. The results are for $L=100$ averaged over $1000$ random realizations of the phase noise (error bars showing the standard deviations).}
 \label{fig3}
\end{figure}

Let us first note a recurrence relation due to the successive application of the unitary matrix $T_k(-h_0)T_k(+h_0)$ (no random phase added)
\begin{align}\label{rec2}
    |v_k(n+1)|^2 &= |c|^2|u_k(n)|^2 + |d|^2|v_k(n)|^2 \nonumber \\
                 &+ [c^*d ~ u_k^*(n) v_k(n) + {\rm c.c.}],    
\end{align}
considering this unitary matrix admits a generic form $\begin{bmatrix}
        a & b \\
        c & d
    \end{bmatrix}$ 
($a, b, c, d$ are complex functions of $k, h_0, T$). In the absence of the cross term in the square parentheses, we readily identify this relation with Eq.~\ref{rec1}, the recurrence relation for the probability count, with $|d|^2\equiv P_{\rm ex}(k)$. The contribution of the interference term is, at this point, explicit in giving rise to DF. 

While the full expressions of $a, b, c, d$ are too complicated to place here, a simplified version can be obtained by inspecting the large $h_0$ limit which also offers additional insights into the peak-valley structure. We find, in this limit, the off-diagonal element of the unitary time-evolution matrix is  
\begin{align}\label{largeh1}
    c \approx -2T e^{i h_0T} \sin(k)\left[\frac{\sin(h_0T)}{h_0 T}\right],
\end{align}
therefore, when $h_0T=2\pi h_0/\omega$ is integer multiples of $\pi$, namely, $2h_0/\omega=p\pi$ -- the freezing condition, the off-diagonal terms of the matrix $T_k(-h_0)T_k(+h_0)$ vanish leading to an arrest of the magnetization dynamics regardless of $k$. As a result, Eq.~\ref{largeh1} renders a peak-valley profile to $m_z(t)$ when plotted against $h_0$ at a fixed frequency of the drive \cite{bhattacharyya2012transverse}. 

When random phases are included by setting $u_k(n)\rightarrow e^{-2i\pi\lambda\phi_k(n)}u_k(n)$ and $v_k(n)\rightarrow e^{2i\pi\lambda\phi_k(n)}v_k(n)$ at the end of each ($n$-th) full cycle, Eq.~\ref{rec2} modifies to
\begin{align}\label{rec3}
    |v_k(n+1)|^2 &= |c|^2|u_k(n)|^2 + |d|^2|v_k(n)|^2 \nonumber \\
                 &+ [c^*d ~ u_k^*(n) v_k(n) e^{4i\pi\lambda\phi_k(n)} + {\rm c.c.}].  
\end{align}
From the results of the large $h_0$ limit, one would then infer that this alters the freezing condition to depend on $k$. More specifically, for a given $p$ satisfying $p\pi=h_0T$ (in the absence of the random phases), a $k$-dependent random shift $\Delta_k$ is introduced for which $p\rightarrow p'=p+\Delta_k$ where $\Delta_k=2\lambda\phi_k$. The parameter $p'$ is not in general an integer anymore (unless $\Delta_k$ is so at some special values of $\lambda$ and $\phi_k$) and the off-diagonal element in Eq.~\ref{largeh1} is also modified to 
\begin{align}\label{largeh2}
    c \approx -2T e^{i (h_0T+2\pi\lambda\phi_k)} \sin(k)\left[\frac{\sin(h_0T)}{h_0 T}\right].
\end{align}
In what follows, we numerically investigate the effect of the resultant interference due to these random phases on the resilience of the overall freezing phenomenon by studying the behavior of $m_z(t)$ and $Q|_{N_cT}$ (after $N_c$ number of full cycles) which are the integrated response of all $k$-modes. We do so by varying the parameter $\lambda$ which, as defined earlier, serves as the effective strength of the random phases $\phi_k(n)$. For the numerical analysis, we fix $L=100$ as no prominent $L$-dependence is observed at larger $L$-values. For $\lambda\neq 0$, an average is performed in computing the response over $1000$ random realizations for each $k$ which turns out to be sufficient in the present context. \\


\begin{figure*}
 \centering
  \includegraphics[width=2.07\columnwidth]{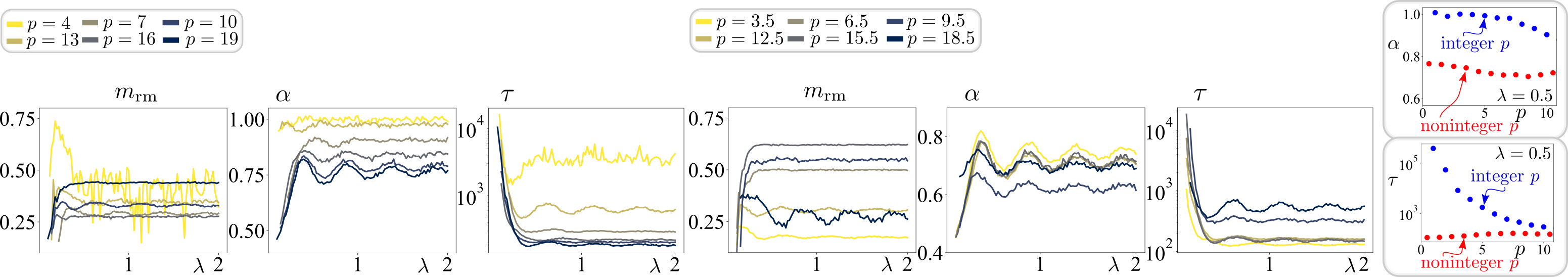}
  \caption{The behavior of the remnant magnetization $m_{\rm rm}$, the steering exponent $\alpha$, and the time scale $\tau$ as a function of the strength of the phase noise $\lambda$ after $N_c=1000$ full cycles (obtained from the fitting of the average $m_z(t)$ over $1000$ random realizations of the phase noise).  The first three figures from the left show results for various integer values of $p$ and the next three show the same for various non-integer values of $p$. The stretched-exponential behavior gains prominence at larger integer values of $p$ or equivalently at slower frequencies, where almost-adiabatic dynamics set in, and approaches the conventional exponential form at higher frequencies (see the right-most figure where the blue-filled circles correspond to the integer values of $p$ or the peaks (freezing points) and the red ones correspond to the half-integers or the valleys). The time-scale $\tau$ is also seen to be significantly higher at the integer values of $p$ (freezing points). The results are for $L=100$.}
 \label{fig4}
\end{figure*}

\begin{figure}
 \centering
  \includegraphics[width=\columnwidth]{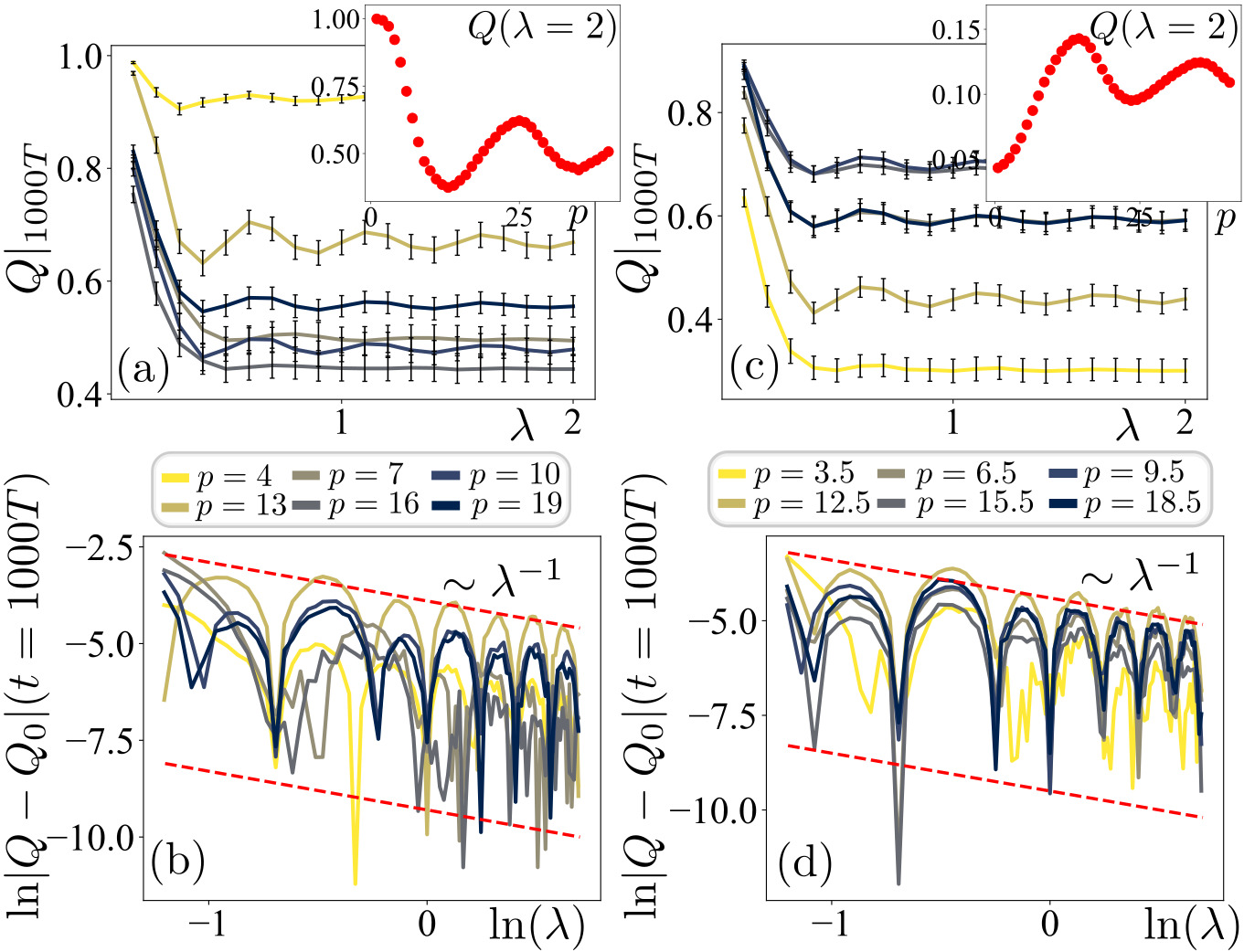}
  \caption{The behavior of $Q|_{1000T}$ (after $1000$ complete cycles) plotted against the effective strength $\lambda$ of the random phases (averaged over $1000$ random realizations). We observe a non-monotonic peak-valley behavior at various frequencies of the periodic drive [here for different values of $p$: integers in {\bf(a)} and nonintegers in {\bf(c)}] with the envelope falling as $\lambda^{-1}$, as shown in {\bf(b)} for the integers and in {\bf(d)} for the nonintegers ($Q_{0}$ denoting the running mean). At a fixed strength of the phase noise $\lambda=2$, the insets display the behavior of $Q|_{1000T}$ vs $p$ for integer values of $p$ in {\bf (a)} and noninteger values of $p$ in {\bf (c)}. The values of $Q|_{1000T}$ at the valleys become closer to those at the peaks with smaller frequencies (or higher $p$ values); the system enters the regime in which almost-adiabatic dynamics dictate. The results are for $L=100$ (error bars showing the standard deviations).}
 \label{fig5}
\end{figure}

\noindent
{\it At the Freezing Peaks:} As seen in Fig.~\ref{fig2}, the qualitative form of the peak-valley structure remains although the difference in the values of $Q|_{N_cT}$ at the peaks and the nearby valleys gets larger with the effective strength $\lambda$ of the random phases. This implies that even when there is random interference between the amplitudes after consecutive cycles, special frequencies do appear at which the deviation of $Q|_{N_cT}$ from the initial magnetization remains almost frozen. The peak locations are well described by the clean case $\lambda=0$. The values of $Q|_{N_cT}$ at these peaks eventually decay with time, however, the rate is much slower compared to that of any noninteger values of $p$. It turns out one can model the decay of $m_z(t)$ for any $\lambda\neq 0$ by a {\it stretched-exponential} form
\begin{align}\label{stretchexpo}
    m_z(t) = m_{\rm rm} + (m_0-m_{\rm rm}) e^{-(t/\tau)^\alpha},
\end{align}
where the remnant magnetization $m_{\rm rm}$ (which can be readily identified with $Q$ in Eq.~\ref{QQ}), the steering exponent $\alpha$, and the time scale $\tau$ all are frequency-dependent, and the initial magnetization, denoted as $m_0$, is set by $h=h_0$ at $t=0$. A non-zero remnant magnetization
indicates a perpetually sustained memory of the initial value of $m^z$ (perpetual persistence of the shards of the emergent conservation law). We demonstrate this in Fig.~\ref{fig3}. For small integer values of $p$, equivalently, for large frequencies of the drive, the profile of $m_z(t)$ is fairly exponential attributed to strong nonadiabaticity in the dynamics. The stretched-exponential form gradually develops at slower frequencies as evident in Fig.~\ref{fig4}. We further find the time-scale ($\tau$) is suppressed with $\lambda$ leading to a faster approach of $m_z(t)$ to its remnant value $m_{\rm rm}$ at the presence of strong phase noise (Fig.~\ref{fig4}) and, for a fixed $\lambda$, it grows with faster frequencies. The parameters here are extracted from an average value of $m^z$ computed over $1000$ random realizations of the phase noise in each cycle. 

Recently, stretched exponential behavior (Kohlrausch law)~\cite{Kohlrausch_Original, StrchdExp_MolElec_Glasses_Review} which is stable, i.e., where the exponent flows away from unity with increasing system size -- has been observed in a large variety of static and Floquet many-body disordered systems when the disorder strength is high \cite{Asmi_Kohlrausch}, while it has been shown to have strong reminiscence in finite systems even in the ergodic disordered phases earlier~\cite{Lezama}. Various pictures with different degrees of complexities have also been proposed to explain this Kohlrausch behaviour \cite{Asmi_Kohlrausch, Anushya_Long, Roderich_Soumya_Modak_Kohlrausch}. 

Here we observe this in a simple translationally invariant system dynamically frozen under a periodic drive. Moreover, the melting here does not obliterate the entire memory of the initial state but leaves behind a steady remnant that depends on the initial conditions. This underscores the stability of the DF states. Our work is inspired by the simplest one where a two-level setting -- a spin-half in a bath made of a disordered spin chain -- captures the phenomenology \cite{Asmi_Kohlrausch}. Our integrable system, following spin-to-fermion transformation and Fourier transform, resolves into a bunch of independent dynamically frozen two-level systems and the random phase noise imprinted independently on each of them gives a stretched-exponential melting of the DF state. 

We also trace the behavior of $Q|_{N_cT}$ by varying the parameter $\lambda$ which features prominent oscillations with a decaying envelope, shown in Fig.~\ref{fig5}. The simple expression in Eq.~\ref{largeh2} at large driving amplitudes is speculative of this feature; because of the modified phase factor, the integrated response is expected to fall as $\lambda^{-1}$ in the leading order: $Q(p)|_{N_cT} = Q_0(p) + A(p)/\lambda$, regardless of $p$. For a fixed $\lambda$, a plot of $Q|_{N_cT}$ as a function $p$ exhibits a similar pattern since the term with $\lambda$ renders an additive correction to the term $h_0T$ as noted in Eq.~\ref{largeh2}.


\noindent
{\it Away From The Freezing Peaks:} Away from the frequencies that respect the freezing condition at $\lambda=0$ maximally ({\it i.e.}, the peaks in Fig.~\ref{fig2}), we observe a prominent stretched-exponential decay of $m_z(t)$ with the time scale $\tau$ much smaller than those at the freezing points or peaks, as evinced from Fig.~\ref{fig3}. Other features of the valleys are qualitatively similar to those at the freezing peaks (cf. Fig.~\ref{fig4}), such as the non-monotonic behavior of $Q|_{N_cT}$ with a decaying envelope as a function of both $\lambda$ and $p$ and the decay of $Q|_{N_cT}$ in the form $\lambda^{-1}$ at large driving amplitudes (Fig.~\ref{fig5}). The values of $Q|_{N_cT}$ at large non-integer values of $p$ approach those at the peaks signaling the freezing to occur at small frequencies even in the presence of interference due to random phase noise, however, the scale is enhanced by the effective strength ($\lambda$) of the noise. \\


\noindent
{\bf Summary and Outlook:} Our investigation focuses on probing the robustness of the phenomena of DF in periodically driven many-body systems against interference from random phase noise. In practical quantum devices designed for exploring out-of-equilibrium quantum dynamics, even the slightest incoherence in the driving protocol can naturally introduce such a noise. In earlier observations, within an Ising chain subjected to coherent periodic driving, specific values of the ratio between the drive frequency $\omega$ and amplitude $h_0$ led to a notable slowing down of transverse magnetization relaxation -- a hallmark known as the freezing condition. This phenomenon manifested as a distinctive peak-valley structure in the late-time response ($Q$) of the system as a function of $\omega$. Our study delves into how random phase accumulation after each drive cycle affects the freezing at these specific frequencies and the overall peak-valley structure. We find that while such random phase interference maintains the qualitative form of the peak-valley structure and the locations of the peaks in the frequency domain, the transverse magnetization $m_z$ eventually decays over time. However, the decay time scale is significantly extended at the freezing peaks compared to frequencies away from these points. Moreover, we observe that increasing the strength ($\lambda$) of the phase noise enhances the disparity in relaxation between $m_z(t)$ at the peaks and the valleys; additionally, a notable {\it stretched-exponential}  decay of $m_z(t)$ emerges as an exclusive feature of the nonadiabatic dynamics (earlier observed in complex systems in the presence of disorder including many-body localization \cite{carmele2015stretched, fischer2016dynamics, levi2016robustness}). Interestingly, as a function of the strength $\lambda$ of the noise, the long-time-averaged response $Q$ exhibits a {\it non-monotonic} behavior, indicating an intricate interplay between the repeated quantum interference and the stochastic process of random phase insertion. Extending our analysis to interacting systems presents a promising avenue for exploration. Recent advancements in cold-atom experiments \cite{lye2005bose, bloch2005ultracold, chen2011controlling} suggest the feasibility of experimentally observing the freezing phenomenon in periodically driven arrays of hard-core bosons in optical lattices. Such experiments hold promise for validating our predictions regarding the stability of many-body DF against the random phase noise considered.\\

\noindent
{\bf Dedication:} We dedicate this paper to our much-loved colleague, friend, and mentor, Late Amit Dutta, who left us all of a sudden in 2022. His untimely demise has left a vacancy in the heart of, we dare say, anyone who came close to him in any capacity. We are no exception.\\

\noindent
{\bf Author contributions:} Arnab Das suggested the problem, analyzed the results, and wrote the manuscript. Krishanu Roychowdhury performed the calculations, prepared the figures, analyzed the results, and wrote the manuscript.\\

\noindent
{\bf Funding:} No additional funding was received particularly to support this work.\\

\noindent
{\bf Data availability statement:} The data may be available on request to the corresponding author.\\

\noindent
{\bf Declarations}

\noindent
{\bf Conflict of interest:} We declare that this work is free of any conflict of interest. The authors have no financial or proprietary interests in any material discussed.


\bibliography{ref}
\end{document}